# Dark-Bright Exciton Splitting Dominates Low-Temperature Diffusion in Halide Perovskite Nanocrystal Assemblies

Andreas J. Bornschlegl[1], Michael F. Lichtenegger*[,1], Leo Luber[1], Carola Lampe[1],

Maryna I. Bodnarchuk[2], Maksym V. Kovalenko[2,3] & Alexander S. Urban*[,1]

*Nanospectroscopy Group and Center for Nanoscience (CeNS), Nano-Institute Munich,*

*Department of Physics, Ludwig-Maximilians-Universität, 80539 Munich, Germany* [1]

*Laboratory for Thin Films and Photovoltaics, Empa–Swiss Federal Laboratories for Materials*

*Science and Technology, 8600 Dübendorf, Switzerland* [2]

*Department of Chemistry and Applied Biosciences, Institute of Inorganic Chemistry, ETH Zürich,*

*8093 Zürich, Switzerland* [3]

**Abstract:**

Semiconductor nanocrystals could replace conventional bulk materials completely in displays and light-emitting diodes. Exciton transport dominates over charge carrier transport for materials with high exciton binding energies and long ligands, such as halide perovskite nanocrystal films. Here, we investigate how beneficial superlattices - nearly perfect 3D nanocrystal assemblies of nanocrystals are to exciton transport. Surprisingly, the high degree of order is not as crucial as the individual nanocrystal size, which strongly influences the splitting of the excitonic manifold into bright and dark states. At very low temperatures, the energetic splitting is large for the smallest nanocrystals, and dark levels with low oscillator strength effectively trap excitons inside individual nanocrystals, suppressing diffusion. The effect is reversed at elevated temperatures, and the larger NC size becomes detrimental to exciton transport due to enhanced exciton trapping and dissociation. Our results reveal that the nanocrystal size must be strongly accounted for in design strategies of future optoelectronic applications.

Since their first conception in the 1980s, semiconductor nanocrystals (NCs), with their size-dependent properties, have been a focus of scientific study as a platform to understand the underlying material in more detail.[1-4] Knowledge so obtained has aided in improving their quality to the degree that they are now being implemented into commercial devices.[5, 6] However, organic ligands are typically required to passivate semiconductor NC surfaces to realize these characteristics and provide colloidal stability.[7, 8] The sizeable inter-NC separation so arising strongly impedes charge transfer. Especially for materials with significant exciton binding energies, exciton transport via Förster resonance energy transfer (FRET) becomes the dominant transport mechanism.[9, 10] Several studies have looked into exciton diffusion in NC thin films, with optical diffusion microscopy as one of the best methods for investigating these processes.[11] It was determined that excitons diffuse via energetic downhill migration, diminishing the diffusivity over time. On a short time scale, migration is enhanced due to the favorable overlap of donor and acceptor FRET pair spectra. However, for long times, excitons become progressively stuck in energetic sinks in a heterogeneous energy landscape, effectively terminating exciton transport. Additionally, decreasing the NC separation (by reducing inorganic shells or using shorter organic ligands) enhanced the diffusivity in the films.[12] One newer class of semiconductor, halide perovskite, offers not only superior optical properties: high photoluminescence quantum yields (PLQYs) and spectral tunability throughout the visible range, but also abundant, less toxic constituents and facile, reproducible, and scalable syntheses.[13-15] Perovskite nanocrystal (PNC) films exhibit far larger diffusivities and diffusion lengths than conventional semiconductor NC films, an effect that has been attributed to a homogeneous energy landscape, high oscillator strengths, large absorption cross sections, and low nonradiative decay rates.[16, 17] However, not all excitons are equal. Quantum confinement to the ultrasmall halide perovskite NCs and only weak screening of the Coulomb interaction of electron and hole through strong dielectric confinement of the surrounding organic ligands lift the degeneracy of the exciton manifold.[18, 19] This, in turn, leads to low-lying "dark" states with weak dipole oscillator strength and higher-lying "bright" excitons with large dipole oscillator strengths.[20-23] The bright-dark splitting depends upon the material, the geometry, and the size of the NCs and can surpass 30 meV, which might affect exciton diffusion even at room temperature.[21, 24-26] To investigate how the energetic structure might affect exciton diffusion, it is imperative to study this process in a temperature-controlled environment on precisely defined and tunable NC systems.[27, 28]

Hence, we here realize a diffusion microscope implemented within a closed-cycle cryostat, enabling a study of exciton diffusion in a temperature range of $9 - 300\ K$. We compare highly ordered 3D assemblies of halide perovskite nanocubes – so-called superlattices (SLs) comprising different NC sizes with drop-cast thin films of a third NC size. Interestingly, the high degree of order in the SLs is not as significant as the actual size of the individual NCs and the temperature of the NC ensemble in determining diffusion efficiencies, which differs from previous results on conventional QD solids.[29] This results from a large quantum- and dielectric-confinement-induced splitting of the excitonic levels into upper "bright" and lower "dark" states in our particular system.[26, 30] The energetic splitting is highly dependent on the size of the NCs as predicted by theoretical models considering short and long interactions between hole and electron states and observed in measurements, which deal with PNC cubes sized from 4 to 10 nm and quasi bulk-like PNCs.[31-33] The dark bright splitting energy is inversely proportional to the NC size, *i.e.*, for smaller NCs the splitting energy goes up and reaches up to 17 meV for 4 nm PNC cube sizes. We estimate a splitting of our smallest PNC cubes of 15 meV, which is equivalent to the thermal energy at 174 K. Accordingly, the effect manifests itself strongly at low temperatures and for the smallest NCs.[24] Optically excited electron-hole pairs rapidly relax to the dark levels, which preclude FRET due to a negligible dipole oscillator strength of the dark excitons. With increasing temperatures, the excitons can be thermally excited into the bright levels, and the dark states only slow down diffusion instead of preventing it. The inter-NC hopping processes are also thermally activated, so hopping rates increase with temperature, further enhancing diffusion. Here, the length of the inter-NC hop is given by the size of the individual NCs. We assume that the center of the exciton wavefunction and the center of the NC coincide due to symmetry considerations. This directly implies that the location of the point dipole, which is an approximation within FRET theory, is also located at the center of the NC. Consequently, an exciton in the 14 nm NCs travels nearly three times the distance compared to the 5 nm NCs. As the temperature increases further, the smaller size becomes beneficial, as weaker binding energies lead to a stronger dissociation of the excitons into free electrons and holes. This, in turn, severely limits diffusion in the larger NC ensembles, as free charge carriers have a greatly suppressed inter-NC transfer component. The results of this study clearly show that it is imperative when designing the ideal NC for optoelectronic applications to consider their size, shape, and order and not only the organic ligand length to enable sufficient transport capabilities.

## Results:

### Basic properties of nanocrystal assemblies

To investigate the effect of NC size and order, we fabricated films by drop-casting 14 nm large cubic PNCs synthesized via tip-sonication (Figure 1a).[34] Optical microscopy and scanning electron microscope (SEM) imaging reveal dense films with no high-range ordering. While there are voids around 50 nm in size, these are few and far between. Up to several tens of PNCs form closed-packed structures, but a long-range order of such structured clusters is missing, *i.e.*, these clusters are randomly orientated to each other. These films sharply contrast superlattices (SL) visible in Figure 1b,c via SEM imaging formed out of 8 nm and 5 nm cubes, respectively. The cubes are synthesized via hot injection and form SLs through solvent-drying-induced spontaneous assembly.[35] The SLs are highly ordered over several microns. No voids or larger/smaller PNCs disturb the SL order. The smaller PNC cubes have lower relative deviations of their lateral size, as observed in transmission electron microscopy (TEM) images (See Figure S1), resulting in a more ordered film formation. All PNCs exhibit high optical quality, as demonstrated by their intense, narrow (full width at half maximum – fwhm $= 75 - 93$ meV) emission spectra (Figure 1d, Table S1). The PL peak blueshifts from 515 nm to 498 nm due to quantum confinement as the PNC size shrinks. The spacing between adjacent PNCs is determined by the organic ligands used to passivate the PNCs and amounts to approximately 2 nm in all films.[36] This separation precludes charge transfer between adjacent NCs, suppressing (macroscopic) electron or hole transport. Thus, the only possible way to transfer electron hole (e-h) pairs through the films, regardless of size and order of the NCs, is via resonant energy transfer.[16]

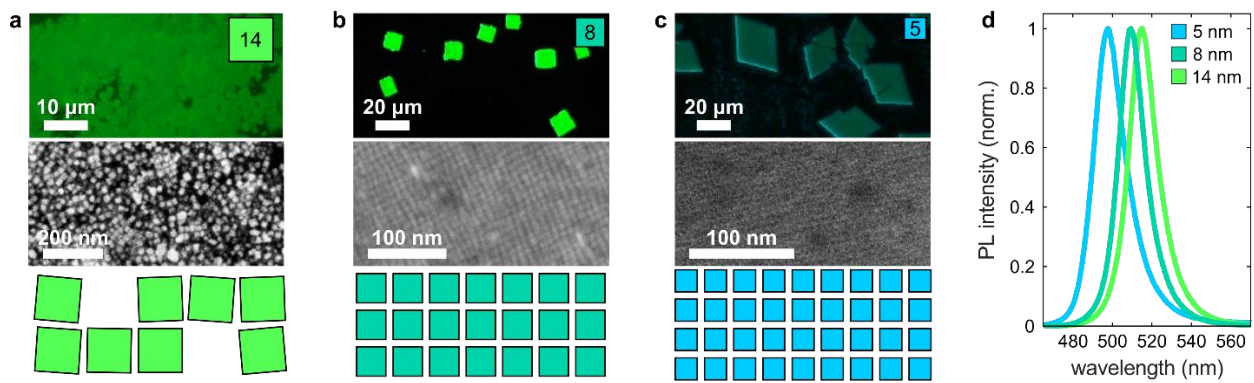

**Figure 1** | PL (upper panel) and SEM (lower panel) image of a **a)** drop-casted 14 nm PNC cube film **b)** SL consisting of 8 nm PNC cubes, and **c)** SL consisting of 5 nm PNC cubes; a schematic of the film quality (order) is positioned below each SEM image illustrating the respective order differences and causal voids. **d)** Room-temperature PL spectra of SL consisting of 5 nm and 8 nm large cubic PNCs and a film of 14 nm large cubic PNCs.

FRET processes are multiplicative and dependent on several conditions, rendering the processes non-efficient if even one of these conditions is weakly fulfilled: 1) The donor photoluminescence

(PL) and acceptor absorption spectra should have a sizable overlap, 2) the orientation of the donor and the acceptor dipole moments should be aligned, *i.e.*, as collinear as possible, and 3) the donor-acceptor distance should be small (not more than 10 nm)[37]. Fortunately, those conditions are highly fulfilled for PNCs, as seen in many preliminary works.[9, 16, 17]

**Low-temperature regime: Bright-dark exciton splitting**

To track exciton diffusion in the PNC structures, we have realized a PL microscope inside a closed-cycle cryostat, yielding spatially and temporally resolved exciton dynamics at temperatures down to 9 K (See Figure S2). Briefly, a laser is focused onto the NC assemblies, exciting electron-hole pairs that rapidly form excitons, which can emit photons and diffuse within the nanostructures. We ensure that we are in a low-power excitation regime, where only few NCs are excited, to preclude non-linear or memory effects from playing a role.[38] Additionally, even though the number of transition dipoles within the excitation spot, the density of states, etc., likely vary between the three samples due to the different ratios of inorganic to organic components, the nature of the physical processes occurring should not vary significantly. A single photon avalanche detector (SPAD) is scanned over the magnified focal plane of the PL, recording a PL decay trace at each spatial position with a temporal resolution of 25 ps (see Figure S3).[11, 12] With temperature affecting many relevant processes for exciton diffusion, such as thermally-induced hopping (see Figure S4), trap formation and binding, and exciton formation and dissociation, this setup can shed light on the nature of the diffusion processes.[11, 39, 40] We carried out multiple heating ramps on the samples to confirm that the samples are not affected permanently by the temperature and the optical measurements are stable over time (see Figure S5). In the measurements, the exciton concentration in an optically excited NC ensemble corresponds directly to the spatial PL emission profile, given by a Voigt function. A broadening of this function is indicative of exciton diffusion. Accordingly, we monitor the change in the PL profile variances $\sigma^2$ (from the Gaussian part of the distribution) in 25 ps steps after excitation, leading to the mean-squared displacement (MSD):

$$MSD(t) = \sigma^2(t) - \sigma^2(0) \qquad (1)$$

We find a striking difference between the smallest and largest NCs at low temperatures. At 9 K, the MSD increases rapidly and nearly linearly within the first one to two nanoseconds for the 5 nm and 8 nm SLs (Figure 2a, see Figure S6). The MSD, however, plateaus and then begins to decline, reaching a value of $MSD = 0.015 \frac{cm^2}{s}$ for the 8 nm SLs and even down to $MSD = 0 \frac{cm^2}{s}$ for the 5 nm SLs after four ns. This latter behavior can be naively interpreted as a negative diffusion. In contrast, for the 14 nm cubes, the MSD values continuously increase over the same time range (see Figure S6). This apparent negative diffusion is also visible at 20 K for the 5 nm SLs; however, the MSD does not return to 0, remaining at a constant value for later times. At 40 K, the negative diffusion disappears, and the MSD increases continuously.

This increase is much faster within the first 500 ps than subsequently. This bimodal behavior becomes less prominent for higher temperatures until the MSD increase becomes monotonous for $T > 80$ K. The initial rapid increase is the same for all these temperatures, suggesting a similar diffusion source. Negative diffusion has been observed previously in different systems. Generally, it requires multi-component systems such as singlet-triplet states in organic semiconductors or intervalley exciton-phonon scattering in 2D transition metal dichalcogenides.[41, 42] This was also seen in 2D Ruddlesden-Popper Perovskites and could be explained due to the existence of two types of excitons: a highly mobile yet short-lived species and a barely mobile and long-lived species.[43] To explain why negative diffusion is prevalent in the 5 nm PNCs but not in the 14 nm PNCs, we must consider the energetic structure of the exciton. Due to the strong quantum and dielectric confinement, the exciton manifold splits into three bright levels and one dark one (Figure 2b). While there had previously been some debate about the ordering of the levels, it has now generally been accepted that the dark exciton is the energetically lowest level.[20, 30] The splitting between the bright and dark levels depends strongly on size. [21, 44] We estimate the splitting energies of the NCs used here to be: $\Delta E_{BD}^{14\,nm} = 3\,meV$; $\Delta E_{BD}^{8\,nm} = 6\,meV$; $\Delta E_{BD}^{5\,nm} = 15\,meV$ (Figure 2c).[21, 24, 31, 45] Using a Boltzmann distribution, we find that at the lowest temperatures, only the 14 nm NCs possess a significant bright-state occupation (see Figure S7). Accordingly, for the smaller NCs, excitons will quickly relax to the dark state, in which they are then fundamentally trapped, while in the larger NCs, excitons will readily switch between bright and dark states. The existence of the dark exciton and the effect of the bright-dark splitting energy can not be seen directly in the PL spectra due to linewidth broadening (see Figure S8), but they can be observed in the time decay of the PL (see Figure S9). For the smallest 5 nm NCs, the decay exhibits two very distinct lifetimes at 10 K, one of approximately $\tau_B = 480\,ps$ and the other $\tau_D = 52\,ns$ (see Figure S10). Importantly, integrating the PL intensity over the two separate time ranges, reveals that roughly 58 % of all photons are emitted from the dark state, while only 42 % stem from the rapid decay of the bright exciton. For the 8 nm NCs, the decay from the dark exciton is also visible at 10 K, yet diminished in comparison to the 5 nm NCs ($\tau_B = 400\,ps$, $\tau_D = 36\,ns$, $I_B = 0.76\,I_{tot}$). For the largest NCs, only the fast decay from the bright excitons can be observed.

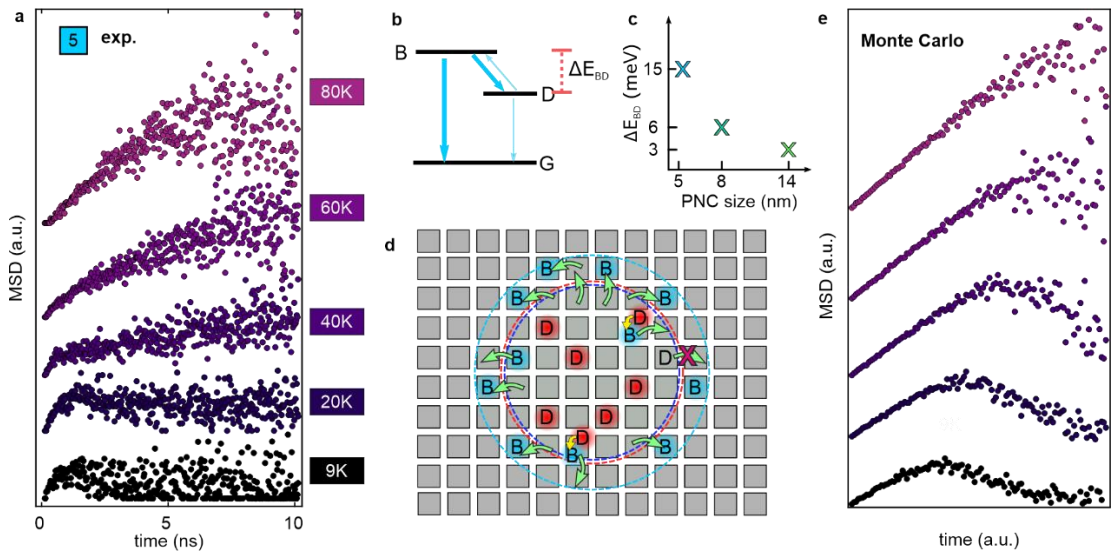

**Figure 2 | a)** MSD curves for temperatures between 9 K and 80 K for the 5 nm SL sample. **b)** Scheme of the exciton fine structure, comprising a dark singlet (D) and a 3-fold degenerate bright triplet state (B). The thickness of the arrows depicts the transition rates. **c)** Bright-dark splitting energies for the three NC systems extracted from literature. [21, 24, 31, 41] **d)** Illustration of the microscale processes underlying the macroscopic exciton diffusion. "B" and "D" denote bright and dark excitons, respectively. Green arrows indicate FRET processes and yellow arrows indicate a thermal excitation of a bright exciton from a dark state. Dark excitons effectively do not undergo FRET processes. The three dashed circles represent the excitation spot of the laser (dark blue), the extent of the dark exciton distribution (red), and the emission stemming from bright excitons (light blue). **e)** MSD results of a Monte Carlo simulation based on the model presented in (b). Here, the dark-to-bright transition rate is successively increased to simulate the system heating up.

The prevalence of the dark exciton is essential when considering that the FRET process, which is likely responsible for the diffusion, is based on dipole-dipole coupling. With the dark states possessing nearly no dipole component (low oscillator strength compared to the bright exciton levels), the probability of a FRET hopping of excitons in the dark state from one exciton to the next is negligible. Thus, only excitons in bright states can undergo FRET processes, and excitons in dark states must first be thermally excited into an upper bright state before they can also diffuse, as indicated in the sketch in Figure 2d by green arrows.

We implemented a Monte-Carlo simulation for this system to confirm whether the dark exciton state can cause the observed negative diffusion (see Methods for details). In this model, the excitons are created nearly equally in bright/dark states. This results from the PNCs being excited far above the exciton levels within the e-h continuum and the charge carriers populating the exciton levels nearly equally upon carrier cooling.[21] An exciton in its bright state can either recombine with a lifetime of approximately 1.5 ns, hop to an adjacent NC, or relax down to a dark state. As previously observed, once inside the dark state, the recombination lifetime increases to 1 µs.[24, 46] Hopping no longer occurs, and the exciton can only be thermally excited to the higher-lying bright states at higher temperatures. The simulated MSD curves match the experimental trends nicely, with a clear negative diffusion trend observable for the lowest temperatures and a bimodal increase for elevated temperatures. The bright excitons can diffuse between NCs, which initiates rapid diffusion. However, these recombine quickly or relax into dark states, which possess lifetimes in the microsecond regime and do not exhibit diffusion. Once all excitons have vacated the bright levels, the distribution returns to the original width, where the dark exciton population is also approximately equal to the size of the exciting laser spot. Accordingly, the apparent negative diffusion is not an actual diffusion process but is caused by the two different exciton species (bright and dark) with drastically different inter-NC transfer rates. At higher temperatures, thermal excitation of the dark excitons becomes progressively more likely, reducing the negative diffusion and inducing the monotonous diffusion for higher temperatures.

For the largest NCs with the smallest bright-dark splitting, the trapping effect into the dark exciton state is not observable and, at most, reduces the overall diffusion.

Importantly, we can rule out that the observed diffusion effects are caused by excitons trapped in defects. Firstly, at the lowest temperatures (down to 9 K), both the 5 nm and 8 nm SLs show a long-lived PL decay, which can be fit by an exponential function, revealing a purely excitonic decay mechanism (see Figures S9, S10). In contrast, the 14 nm NCs show a power-law decay, which is indicative of trapping and a delayed release of excitons at defects (see Figure S11).[47-49] The SLs show similar behavior for elevated temperatures, where the dark excitons can be thermally excited into the bright states. Moreover, trap-mediated recombination generally increases with temperature, which contradicts our findings of enhanced diffusion in the temperature range of 9 - 100 K.[50]

We can also rule out the presence of ultrafast exciton transport of delocalized excitons or inter-NC electronic coupling due to the ratio of exciton Bohr radius to NC radius $\left(\text{here: } \frac{r_B}{r_{QD}} = 0.5 - 1.4\right)$ being significantly smaller than necessary to enter this regime $(r_B \gg r_{QD})$[51] and the PL profile of the first frame (at 25 ps) being independent of temperature for all samples (see Figure S12). Superfluorescence was not observable within the applied excitation power regime. This is most likely due to the exciton density in the TRPL diffusion measurements having to be reduced sufficiently to prevent exciton-exciton interaction and allow unperturbed energy transfer.[52, 53]

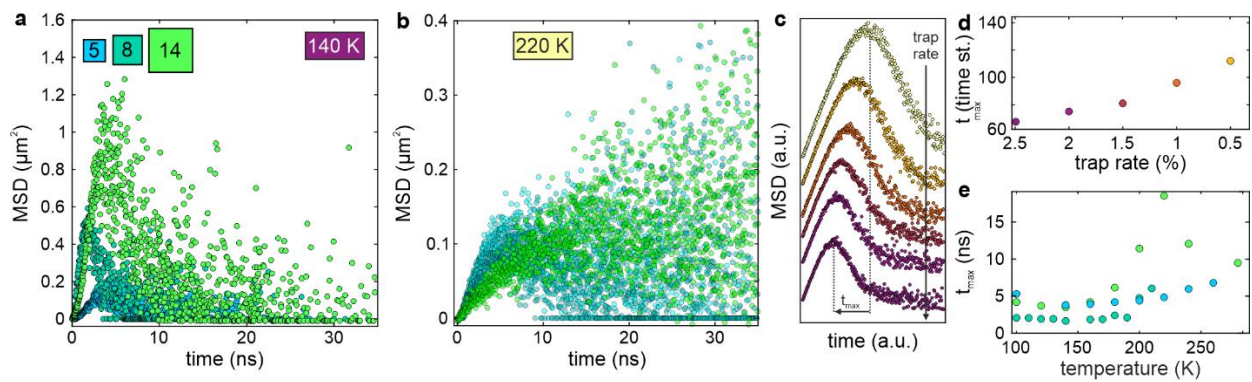

**Figure 3 |** MSD curves for all NC species at **a)** 140 K and **b)** 220 K. **c)** Monte-Carlo-simulated MSD curves employing the trap model introduced in the main text. The exciton trap rate, which is inversely connected to the trap density, increases from bottom to top (see Figure S13 for details on the model). d) By decreasing the trap rate, the MSD maximum shifts to earlier times. e) $t_{max}$ values from the MSD curves from the experiment for all temperatures and all NC species.

**High-temperature regime: Trap states and exciton dissociation**

As the temperature increases further, the MSD curves of all NC samples exhibit a steeper increase, signifying an enhanced diffusion. Three effects can explain this: 1) Homogeneous broadening increases leading to a more significant spectral overlap and an enhanced FRET rate,[54] 2) the increased exciton lifetime allows the excitons to diffuse for a longer time than at low temperatures, and 3) thermally activated hopping is more efficient at higher T, reducing the detrimental effect of an inhomogeneous energetic landscape (see Figure S4).[12, 55] The observed effect progresses up to 100 – 150 K, upon which we again observe a contraction of the MSD (Figure 3a,b), albeit for much longer times on the order of 10 – 30 ns. Moreover, the time at which the contraction begins, $t_{max}$, is highly temperature dependent, with higher temperatures leading to larger values. The bright-dark exciton state cannot explain this observation, as at these temperatures, excitons do not reside long in the dark state, even for the smallest PNCs. As explained in the previous section, we could exclude the contribution from excitons binding to trap states at low temperatures, however can no longer do this for elevated temperatures, where binding of excitons and charge carriers to defects can become highly detrimental to optoelectronic properties.[56] Accordingly, we reworked our Monte-Carlo simulation to reproduce the observed MSD traces. In this new model, we introduced a trapping rate of free excitons into deep defects from which the excitons can still emit (albeit with a radiative rate of 1/100th that of the free exciton) but not be detrapped (see Figure S13). The result is the bottom curve in Figure 3c, which produces a near-linear increase in MSD up to a maximum value and time (denoted $t_{max}$), subsequently decaying back to zero. An additional contribution due to shallow traps from which the excitons can be detrapped, would only reduce the overall diffusion (analogous to the previous bright-dark model with thermal excitation) and not shift the $t_{max}$ value and is thus omitted from the new model. The increase in the initial diffusion occurring for higher temperatures can easily be reproduced by varying the inter-NC hopping rate, but this does not affect the $t_{max}$ value (see Figure S14). The only way to shift this value to later/earlier times is through a variable trap density, which we incorporate by modifying the trapping rate of the free excitons (Figure 3c,d). Notably, a lower trapping rate (corresponding to a lower trap density) shifts $t_{max}$ to higher values, which agrees nicely with the experimentally observed trends (Figure 3e). Interestingly, we observe essentially constant $t_{max}$ values up to a specific temperature (14 nm: 140 K, 8 nm: 180 K, 5 nm: 220 K), after which the values increase abruptly. This result implies that the number of trap states from which the excitons cannot be detrapped is reduced as the temperatures increase. This assumption can be explained by considering that the traps in the NCs exhibit a breadth of energies that are all deep enough to preclude detrapping at lower temperatures. A certain fraction can no longer be considered deep enough at higher temperatures as excitons can acquire enough thermal energy to detrap and continue diffusing. This reduces the effective deep trap density. The trap depth seems to vary with NC size, either due to the exciton Bohr radius or the binding energy, which are inversely proportional to each other, or the larger surface-to-volume ratio for smaller NCs, since traps are more likely to be located at the NC boundaries.

**Temperature dependence of diffusion parameters**

Understanding these processes, we evaluated the diffusion for all PNC samples and temperatures. Depending on the system and the experimental circumstances, various methods can be employed to reproduce the observed MSD curves, including a subdiffusion model[57] and trapping models of varying complexity.[57-59] While each model can be applied to specific combinations of PNC samples and temperature ranges, none of these models can reproduce the complete set of MSD curves of all three PNC types. However, none of the previously reported data suggests that diffusion should occur differently for the three systems. Accordingly, we numerically determine the diffusivity for all systems and temperatures with the following equation:

$$D_{avg} = \frac{1}{\sum_i^N I_{PL}(t_i)} \sum_i^N \frac{MSD(t_{i+1}) - MSD(t_i)}{2(t_{i+1} - t_i)} I_{PL}(t_i) \qquad (2)$$

The average diffusivity is given by summing the differences in MSD from two subsequent time bins (separated by $t_{bin} = t_{i+1} - t_i = 25$ ps), which are weighted by the total PL intensity at the first time bin up to the time bin corresponding to five times the PL lifetime ($N = 5 * \tau_{PL}/t_{bin}$). The PL lifetimes differ for each NC system and temperature. The total PL intensity $I_{PL}(t_i)$ is the sum of the PL counts of all captured PL decay traces for each time bin, and the PL lifetime corresponds to the time after which total PL intensity has dropped to 1/e of the initial value. This way, we can account for the exciton population, *i.e.*, more excitons are present in early times. Thus, those times should have a more considerable impact on the diffusivity compared to later times, where the exciton population rapidly decreases due to radiative and nonradiative recombination. If normal diffusion is present, *i.e.,* a linear MSD behavior with time, its slope is constant, and Equation 2 expresses the normal diffusivity: $D_{normal} = \frac{1}{2}\frac{\Delta MSD}{\Delta t}$. Importantly, we assure that even at these very late times, when most of the exciton population has decayed, their spatial distribution can still be given by a Voigt function (see Figure S15). For a direct comparison between different temperatures and NC ensembles, Figure S16 shows the PL dynamics during the first six nanoseconds after excitation. With the so-obtained diffusivity and the PL lifetime measured through time-correlated single photon counting (TCSPC), we can then calculate the diffusion length $L = \sqrt{\tau_{PL} D_{avg}}$. All three of these quantities are depicted in Figure 4a-c.

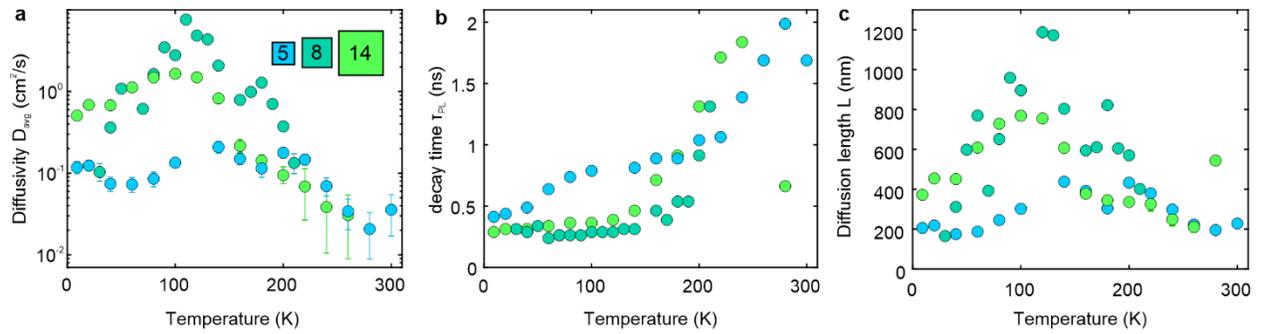

**Figure 4 |** Diffusion characteristics of the NC systems at temperatures ranging from 9 K to 300 K: **a)** Exciton diffusivities calculated with Eq. 2. **b)** PL decay times obtained through TCSPC measurements, and **c)** diffusion lengths calculated from the two previous parameters.

At the lowest temperatures, the diffusivity of the 14 nm NC films is largest at $D_{avg} = 5 \cdot 10^{-1} \frac{cm^2}{s}$ with both the 5 $nm$ and 8 $nm$ superlattices exhibiting substantially lower values (Figure 4a). This is most likely due to the prevalence of the dark exciton, which severely limits diffusion in the SLs (see Fig 2b,c). Also, considering a FRET-mediated diffusion process between equally spaced NCs (given by the ligand length), the distance covered by a single hopping process is given by the size of the respective NCs plus the ligand separation. This value is nearly three times as large for the 14 nm NCs as for the 5 nm NCs. Accordingly, an exciton would need three successive hops to cover the same distance in the 5 nm SLs. As the temperature is increased, the diffusivity in the larger two samples increases, albeit at different rates, due to enhanced hopping rates (larger FRET rates and thermally activated hopping bridging energy inhomogeneities between adjacent NCs, as shown in Figure S4).[12, 55] Especially for the 8 nm SLs, a significantly increased thermal excitation of the excitons from dark to bright states can be observed, leading to the diffusivity in the 8 nm SLs surpassing that of the 14 nm thin films and reaching a maximum value of $D_{avg} = 7.5 \frac{cm^2}{s}$ at 120 K. Here, the high degree of order in the SL becomes essential and can compensate for the smaller hopping distances. In contrast to these two samples, the diffusivity in the 5 nm SLs fluctuates around the low-temperature value, with no discernible temperature dependence. In this intermediate temperature range, the NC size and the order in the ensembles play similarly crucial roles in determining the diffusivities.

All samples show a steep decline in diffusivity for higher temperatures. To explain this, we again consider exciton traps in the NCs. Shallow traps likely increase in number with rising temperatures as defects are thermally ionized. Moreover, deep traps become progressively shallower with respect to thermal energy, leading to a transient binding, which reduces the overall diffusivity. The importance of bound exciton emission in perovskites was previously shown in organic/inorganic 2D and 3D halide perovskite films.[60, 61] This assertion also matches the increasing $t_{max}$ values due to a lower deep trap density observed in the MSD measurements and Monte Carlo simulations (Figure 3d,e).

At $220\,K$, all three diffusivities become comparable. For higher temperatures, the 5 nm SLs exhibit the highest values (values for the 8nm SLs could not be determined due to sample degradation in this intermediate regime), which are comparable to those in our previous report on PNCs at temperatures around RT.[62] This continued decline coincides with a steep rise in the PL lifetimes (Figure 4b) and a significant drop-off in the total PL count rate at those temperatures (see Figure S17). Notably, the temperature dependence of the PNC samples matches nicely with previous studies on CdSe quantum dots, which identified specific temperature-dependent effects, strongly impacting the overall PL intensity.[63] However, the underlying effects must differ as these two processes begin at higher temperatures than the diffusivity decline. At elevated temperatures, the thermal energy available noticeably affects the Coulomb interaction of electron and hole, in turn influencing the exciton to free charge carrier ratio in the NCs, as given by the Saha equation. This equation quantifies how the free charge carrier density increases with rising temperatures as excitons become progressively dissociated, especially for the less-confined NCs in the intermediate temperature ranges (See Figure S18). Inter-NC transport of free charge carriers is prohibited, further reducing diffusivities. Additionally, the charged particles are more susceptible to binding to trap states, so in the more polarizable larger NCs, the diffusivity is even more severely reduced. This residual discrepancy in the absolute diffusivity values to our previous publication is likely a result of the experimental setup. With the temperature measured just below the substrate but diffusion measurements carried out on top of the NC ensembles, which are hundreds of nanometers thick, the actual temperature in the NC samples is likely higher than indicated by the setup. We corroborate this through additional temperature-dependent PL measurements, which suggest that the temperature in the relevant part of the NC ensembles is significantly higher than anticipated (see Supporting Information, Figures S19, S20). We calculate the exciton diffusion lengths using the obtained values for the diffusivity and PL lifetime (Figure 4c). At room temperature, the values coincide nicely with previously determined values in comparable systems.[12, 57, 62, 64] However, at lower temperatures, the diffusion lengths increase drastically, reaching values of 750 nm for the 14 nm NCs (at 100 K) and nearly 1.2 μm for the 8 nm SLs at 120 K. To our knowledge, these are the highest reported values in nanocrystalline systems, and show the high degree of homogeneity and order in the SL systems.

**Discussion**

The diffusion in colloidal NC thin films and SLs is far more complicated than previously assumed, depending considerably on the observed temperature range. Naively, one would assume that the extremely high ordering and near monodisperse sizes in the SLs dramatically increase diffusion due to enhanced FRET rates, reduced trap densities, or potentially electronic-coupling mediated super-diffusion. Some of these effects likely play a (minor) role, with higher diffusivities in ordered versus nonordered systems. However, several other effects assume more prominent roles when comparing systems comprising NCs of different sizes: The interplay of NC order, individual NC size, and temperature must be carefully considered concerning exciton diffusion performance. At the lowest temperatures, the splitting of excitons into bright and dark states becomes the dominant effect. The dark excitons possess a weak dipole oscillator strength, a prerequisite for FRET processes. As the bright-dark splitting is most prominent in the smallest NCs, excitons relax rapidly to dark states, in which they are stuck energetically and spatially, leading to minuscule diffusivities which are only caused by an initial population of excitons in bright states diffusing until they recombine or themselves become trapped in dark states. Consequently, a larger NC size benefits exciton diffusion in this temperature regime.

At intermediate temperatures (up to approximately 140 K), where dark excitons can be thermally excited into bright states and FRET processes become prevalent, the larger NCs still show far higher diffusivities, as each hopping event brings about a spatial displacement on the order of the NC size. Accordingly, an exciton must hop thrice in the 5 nm SL to diffuse the same distance as an exciton in the 14 nm NC film. However, in this regime, the 8 nm SLs overtake the disordered 14 nm NC films in diffusivity. This trend becomes even more prominent at higher temperatures, especially for the smallest 5 nm NCs. Again, this is an effect of size, as exciton dissociation into free charge carriers, which cannot diffuse due to the sizeable inter-NC separation and trapping effects, becomes dominant, deterring diffusion far less in the smaller NCs. Accordingly, in the intermediate temperature regime, the 8 nm NCs' diffusivity approaches $10\ cm^2/s$, and the diffusion lengths surpass $1\ \mu m$, to our knowledge, the largest value reported in NC thin film systems.[38] While we cannot rule out that defects or other mechanisms play a role in determining the diffusion characteristics of PNC assemblies, the acquired data, in combination with Monte-Carlo modeling, strongly suggest that the exciton fine-structure and binding energies play the predominant role. These results highlight the importance of considering NC morphology, size, and spacing when devising optoelectronic device architectures that rely on efficient carrier transport, such as LEDs or solar cells.

## Methods

**Nanocrystal Synthesis**

<u>14 nm PNCs:</u> The 14 nm PNCs were synthesized as previously described.[34]

Materials: $Cs_2CO_3$ (Cesium carbonate, 99%), $PbCl_2$ (lead(II) chloride, 98%), $PbBr_2$ (lead(II) bromide 98%), $PbI_2$ (lead(II) iodide 99%), mineral oil (light), $SnBr_2$ (Tin(II) bromide), 1-octadecene (technical grade 90%), oleic acid (technical grade 90%), oleylamine (technical grade 70%), hexane (HPLC, grade ≥97.0%, GC)and trioctylphosphine (97%) were purchased from Sigma-Aldrich.

Synthesis: 10 ml octadecene together with 0.5 ml oleic acid and 0.5 ml oleylamine were added to $Cs_2CO_3$ (0.1 mmol) and $PbBr_2$ (0.3 mmol) precursor powders. Then, the reaction medium was subjected to tip-sonication (Sonoplus hd 3100, Bandelin) at a power of 30 W for 10 minutes. The as-prepared NC dispersions were purified by centrifugation at a speed of 9000 rpm for 10 min to remove unreacted precursors, and then the NC precipitates were redispersed in 5 ml hexane under mild sonication. The obtained NC dispersions were centrifuged again at 2000 rpm, to remove large NCs.

<u>8 nm PNC:</u> The 8 nm PNCs were synthesized as previously described.[65]

Stock solutions: $PbBr_2$ stock solution (0.04 M) is prepared by dissolving $PbBr_2$ (1 mmol, 0.367 g, Sigma Aldrich 99.999%) and TOPO (5 mmol, 2.15 g, Strem, 90 %) in octane (5 mL) at 100 °C, followed by dilution with hexane (20 mL). The CsDOPA solution (0.02 M) is prepared by mixing $Cs_2CO_3$ (100 mg, Sigma Aldrich) with diisooctylphosphinic acid (1 mL, Sigma Aldrich) and octane (2 mL) at 100 °C, followed by dilution with hexane (27 mL).

Synthesis: In a 25 mL one-neck round bottom flask, 0.7 mL $PbBr_2$ stock solution is combined with 6 mL hexane. Under vigorous stirring, 0.35 mL of Cs-DOPA stock solution is swiftly injected. After 4 min, 75 µL of didodecyldimethylamonium bromide (DDAB) in toluene (300 mg of DDAB from Alfa Aesar are solubilized in 3 mL anhydrous toluene) is added, and then the reaction is stopped in 2 min.

Purification: The volume of the reaction mixture is decreased to 1 mL by evaporation on the rotary evaporator at room temperature. A ca. 3-fold excess of antisolvent (ethyl acetate and acetone mixture, v/v 2:1) is added, and NCs are isolated by centrifugation at 12100 g for 2 minutes and redispersed in 1 mL anhydrous toluene.

<u>5 nm PNC:</u> The 5 nm PNCs were synthesized by adapting a synthesis procedure.[66]

Stock solutions: Cesium oleate 0.16 M in octadecene (ODE): $Cs_2CO_3$ (250 mg, 0.77 mmol, Sigma Aldrich), and oleic acid (0.8 mL), and octadecene (8.8 mL) are mixed in a 25 mL flask. The mixture is degassed three times and then heated to 100 - 120 °C under $N_2$ until it becomes clear. Cesium oleate in ODE is stored in the glove box.

Synthesis: $PbBr_2$ (75 mg, 0.2 mmol, Sigma Aldrich, 99.999%), $ZnBr_2$ (Alfa Aesar, 180 mg, 0.8 mmol), and distilled mesitylene (5 mL, Acros) are mixed in a 25 mL flask under $N_2$, stirring at 1400 rpm. The mixture is heated to 120 °C: distilled oleylamine (2 mL) and dried oleic acid (2 mL) are injected. The mixture is heated to 145 °C: 0.4 mL cesium oleate is injected from a 0.5 mL glass syringe. The reaction is quenched after 15 $s$ with an ice bath.

Size selection and purification: The crude solution is centrifuged for 3 min at 20130 g, and the precipitate is discarded. 27 mL ethyl acetate is added to the supernatant, then centrifuged for 5 min at 20130 g, and the supernatant is discarded. The precipitate is dispersed in 1 mL of anhydrous toluene.

Post-synthesis treatment: 1 mL NCs in toluene is mixed with 100 µL 0.01 M DDAB in toluene and stirred for one h. 0.01 M solution of DDAB in toluene is prepared by dissolving 9.2 mg DDAB in 2 mL anhydrous toluene.

**Self-assembly of CsPbBr$_3$ NCs**

The 14 nm PNC film was prepared by drop casting 40 µL of the PNC dispersion onto a SiO$_2$-coated (300 nm) 10 × 10 mm Si substrate.

The SLs were prepared on square 5 mm × 5 mm silicon substrates. Shortly before self-assembly, the silicon substrate was dipped into a 4 % HF in water for 20 min, followed by intensive washing with deionized water. The substrate was placed in a 10 mm × 10 mm × 10 mm Teflon well, and 7 µl of stock solution of NCs in toluene (5 nm PNC: 20 µL of NC stock solution in 0.16 mL anhydrous toluene/8 nm PNC: 25 µL of NC stock solution in 0.15 mL anhydrous toluene) were spread onto the substrate. The well was covered with a glass slide to allow slow solvent evaporation. 3D SLs of CsPbBr$_3$ NCs were formed upon complete evaporation of the toluene.

**Confocal low-temperature microscopy**

The setup for conducting exciton diffusion measurements (via TCPSC measurements) and capturing steady-state PL spectra is depicted in Figure S2.

The excitation is provided by a pulsed laser (NKT Photonics, SuperK Fianium FIU-15) and a tunable multiline filter (NKT Photonics, SuperK VARIA) to select an excitation wavelength depending on the specific NC according to SI Table 2. A motorized continuous neutral density (ND) filter wheel (Thorlabs, NDM4) attenuates the signal by reflecting a tunable beam fraction. Together with a power diode (Thorlabs, S120C), this allows for precise control over the irradiance of the laser spot on the sample (SI Table 2). The laser beam is reflected in the direction of the focusing optics inside the cryostat with the help of a suitable dichroic (DC) mirror (either Semrock, HC 495 or Chroma, T 450 LPXR; see SI Table 2). The insides of the closed-cycle helium-cooled cryostat (attocube, attoDRY800) consist of a vacuum objective (Zeiss, EC Epiplan-Neofluar 100x/0.9 NA DIC Vac M27) with a working distance of 1.0 mm and a sample holder on a 3-axis piezo-controlled translation stage (attocube). A thermal bridge between the cold plate and the sample holder allows for sample temperatures from 9 K to 300 K. The generated PL is collected and collimated by the vacuum objective and exits the cryostat in the direction of the DC mirror. The DC mirror transmits the PL signal, thus separating it from the excitation signal (laser beam). An additional suitable long pass (LP) filter (Semrock, 496/LP-25 or Semrock, AT465lp; see SI Table 2) ensures no laser light is left in the signal path by reflecting all light below its characteristic wavelength. A beam splitter (Thorlabs, BSN10R) guides 10% of the PL signal to the (i) spectrometer and 90% to the (ii) TCSPC system. (i) A lens ($f$ = 200 mm) focuses the PL signal into the slit of the spectrometer (Teledyne Princeton Instruments, HRS-500-MS). Inside the spectrometer, the PL is reflected from different optical parts on a rotating turret and detected by a charge-coupled device camera (Teledyne Princeton Instruments, PIXIS 400BR eXcelon). PL images or PL spectra can be recorded by setting the turret to a mirror or a grating. (ii) A lens ($f$ = 200 mm) focuses the PL light onto a plane, where it is magnified by a factor of 122 compared to the initial sample plane (same for (i)). The magnification factor was determined with a gold scale (see Figure S21). The FWHM of the detection point spread function is 270 nm, very similar to those in comparable setups (see Figure S22).[11] A motorized ND filter wheel (Thorlabs, NDM4) can attenuate the PL signal, e.g., to guarantee single-photon statistics in TCSPC measurements. The PL signal is collected by a $10 \ \mu m$ wide glass fiber (ThorLabs, M64L01) connected to a fiber-coupled SPAD (MPD, PDM series). Both laser and SPAD are connected to a TCSPC card (PicoQuant, TimeHarp 260) in the PC to allow for TCSPC measurements. The glass fiber is mounted on a piezo-controlled 2-axis translation stage (Physik Instrumente, Q-521), which enables capturing local PL decays at different positions of the magnified PL signal. Exciton diffusion is measured using this method, as depicted in Figure S3.

**Additional microscopy methods**

Optical images of the superlattices were obtained using an optical microscope Leica DM4 M under UV light.

Transmission electron microscopy (TEM) images of the 14 nm PNC samples were recorded using a JEOL JEM-1011 with an accelerating voltage of 80 kV. TEM images for the $5 \ nm$ and $8 \ nm$ PNC samples were collected using a JEOL JEM2200FS microscope operating at 200 kV accelerating voltage.

Scanning electron microscopy (SEM) images of all samples were recorded using a ZEISS SEM Ultra Plus with an acceleration voltage of 3 kV and an FEI Helios 660 operated at 3 kV using immersion mode.

**Characterization of Exciton Diffusion**

We fitted Voigt profiles to each spatial PL intensity distribution for every time bin to characterize exciton diffusion. A Voigt profile, being the convolution of a Gaussian and a Lorentzian function, can be expressed as an exciton density according to:

$$n(x,t) = e^{-\frac{t}{\tau}}[L(x,0) * G(x,t)] = e^{-\frac{t}{\tau}} V[\mu, A, \sigma, \lambda](x,t).$$

Here, $\tau$ is the monomolecular recombination time, and $A$ is the profile height. For the first time bin, both broadening parameters of the Lorentzian $\lambda$ and the Gaussian part $\sigma$ serve as free fit parameters for the Voigt profile. For all subsequent time bins, only the parameters of the Gaussian part act as free fit parameters. The Lorentzian parameters retain their initial values since the Lorentzian profile best describes the starting exciton distribution. Thus, it does not evolve, implying that the time-dependent Gaussian line shape is the only parameter characterizing the PL broadening/exciton diffusion. The diffusivity $D_{avg}$ is numerically determined with equation (2) in the main text. The diffusion length is calculated with the equation $L = \sqrt{\tau_{PL} D_{avg}}$, where $\tau_{PL}$ is the $1/e$-decay time of the total PL decay trace (integration of PL decay traces at all positions over time).

**Monte Carlo Simulations**

The structure formed by NCs in the experimental setup is implemented as a regular 2D grid without effective boundaries and comprising uniform squares with identical orientations. We found that the results for 2D and 3D grids were nearly identical; however, the computation time in 3D was significantly higher. We, therefore, chose to adopt the 2D system for the model. To prevent the exciton dynamics from being limited by the edges of the simulated environment, sufficiently large dimensions are chosen (dim = 150 lattice units). The side length of a square corresponds to the average center-to-center distance of the respective NCs in the experiment. This distance is also the step size considered for nonradiative transitions between the two closest neighbors within the grid.

The lattice is initially populated with excitons residing in bright, dark, and trapped states, with the ratio depending on the temperature regime. The spatial distribution of the excitons is given by an initial Gaussian function with standard deviation $\sigma = 10$ in lattice units, i.e., 170 nm for the largest NCs considered. The initial excitation marks the starting point of a Monte-Carlo-based random walk for 200-300 frames (after that time, no excitons are left for the considered parameters). The system's dynamics are obtained by considering all possible interactions, such as radiative and nonradiative decay, dissociation, trapping, and inter-NC hopping, assigning each a probability and determining what happens to each exciton for each time step.

We set the parameters of the Monte-Carlo simulations so that they would qualitatively reproduce the MSD data we obtained. Beginning at the lowest temperature for the 5 nm SLs, we set the PL decay rate to reproduce the experimentally determined decay rate (approx. $\tau_{PL} =$

0.8 $ns$), yielding a probability of $\gamma_B = 0.03$. Accordingly, for every time step (corresponding to 25 ps to match the experimental frame rate), there is a 3% chance of a radiative decay of the exciton. The decay rate of the dark exciton was set to match the 100 times longer decay: $\gamma_D = 0.0003$.[21] Nonradiative decay did not significantly affect the results of the Monte-Carlo simulation and was left out of the model. 'Decay' events correspond to the end of the random walk, and the location of the decay event is registered to create the MSD(t) function.

To match the MSD curves, we have two free parameters: the rate with which the bright exciton can relax to a dark state $\gamma_{BD}$ and the hopping rate: $\gamma_{hop}$. For the latter, we assume a FRET rate and consider the number of nearest NCs and next nearest NCs (each four) and the respective distances so that: $\gamma_{hop} = 4 \cdot \left(\frac{1}{1^6} + \frac{1}{2^6}\right) \cdot \gamma_{FRET}$. Hopping events for excitons in a dark or trapped state are negligible due to the weak oscillator strength and weak spatial overlap, respectively.

The best fit to the 5 nm diffusion data at 9 K was obtained for $\gamma_{FRET} = 0.09$ and $\gamma_{BD} = 0.15$. The reverse transition of the exciton from the dark to the bright state does not occur at the lowest temperatures; hence, $\gamma_{DB}(T = 9K) = 0.0$. This value is critical; otherwise, the MSD curves will not return to $MSD = 0$. For higher temperatures, this transition becomes possible, and so we vary the rate: $\gamma_{DB}(T > 9K) = 0.0 - 0.1$, while keeping $\gamma_{BD} = 0.15$ constant. We find that the FRET hopping rate also increases with temperature and assumes values of $\gamma_{FRET} = 0.09 - 0.15$ (See Figure S13). The transition from a bright state to a trap state is also temperature-dependent and varies from 0 to 0.025; detrapping is not permitted, as explained in the main text. In all, we repeated the simulations beginning with the initial Gaussian distributions approximately

10.000 times to obtain an accurate picture of the diffusion behavior of excitons in such NC environments.

For the other two samples, we proceeded similarly, except that we already included the dark-to-bright transition at the lowest temperatures (14nm: 9K, 8nm: 30K) and matched the FRET and the dark-to-bright transition rates, $\gamma_{FRET}$ and $\gamma_{DB}$, to match the experimentally determined curves.

**Data availability**

The authors declare that the data needed to evaluate the conclusions in this manuscript are present in the main text or the Supplementary Information. Experimental procedures, characterization of materials, computational details, Supplementary Tables, and Supplementary Figures are available in the Supplementary Information. Additional raw data formats are available upon request to the corresponding authors.

**Code availability**

The computer simulation codes used in this study are available upon reasonable request.

## Acknowledgments

This project was funded by the European Research Council Horizon 2020 through the ERC Grant Agreement PINNACLE (759744), by the Deutsche Forschungsgemeinsschaft (DFG) under Germany's Excellence Strategy EXC 2089/1-390776260 and by the Bavarian State Ministry of Science, Research and Arts through the grant "Solar Technologies go Hybrid (SolTech)".

## Author Contributions

A.J.B. and M.F.L. contributed equally to this work. A.J.B. and M.F.L designed the study. M.F.L. led the experimental work and processing of experimental data. A.J.B. set up the diffusion measurement technique, aided by M.F.L. L.L. and M.F.L. performed the Monte Carlo simulations. C.L. and M.I.B. prepared the perovskite materials. A.S.U. and M.V.K. supervised the project. A.J.B., M.F.L., and A.S.U. wrote the original draft of the paper.

## Competing Interests

The authors declare no competing interests.

## Additional Information

Supplementary information is online

## Table of Contents

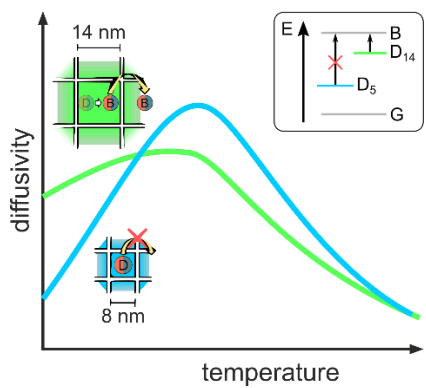

At cryogenic temperatures, bright-dark exciton splitting leads to immobile dark excitons in halide perovskite nanocrystal assemblies, substantially limiting diffusion, especially in strongly confined nanocrystals. At elevated temperatures, this is reversed as thermal excitation into the bright states becomes possible, hopping rates increase, and excitons dissociate into free electron-hole pairs, especially in the least confined nanocrystals.